\documentclass[12pt,draftcls,onecolumn]{IEEEtran}
\usepackage{setspace}
\usepackage{amsmath,amsfonts}
\usepackage{algorithmic}
\usepackage{algorithm}
\usepackage{array}
\usepackage[caption=false,font=normalsize,labelfont=sf,textfont=sf]{subfig}
\usepackage{textcomp}
\usepackage{stfloats}
\usepackage{url}
\usepackage{verbatim}
\usepackage{xcolor}
\usepackage{graphicx}
\usepackage{cite}

\begin{document}

\title{Stacked Intelligent Metasurfaces for \\Integrated Sensing and Communications}

\author{Haoxian Niu, Jiancheng An,~\IEEEmembership{Member,~IEEE},\\Anastasios Papazafeiropoulos,~\IEEEmembership{Senior Member,~IEEE}, Lu Gan,~\IEEEmembership{Member,~IEEE}, Symeon Chatzinotas,~\IEEEmembership{Fellow,~IEEE}, and M\'erouane Debbah, \emph{Fellow, IEEE}
\vspace{-1cm}

\thanks{This work was partially supported by Sichuan Science and Technology Program under Grant 2023YFSY0008 and 2023YFG0291. H. Niu and L. Gan are with the national key laboratory on blind signal processing, School of Information and Communication Engineering, University of Electronic Science and Technology of China (UESTC), Chengdu, Sichuan, 611731, China. L. Gan is also with the Yibin Institute of UESTC, Yibin 644000, China (e-mail: haoxian\_niu@163.com; ganlu@uestc.edu.cn). J. An is with the School of Electrical and Electronics Engineering, Nanyang Technological University, Singapore 639798 (e-mail: jiancheng\_an@163.com). 
A. Papazafeiropoulos is with the Communications and Intelligent Systems Research Group, University of Hertfordshire, Hatfield AL10 9AB, U. K., and with SnT at the University of Luxembourg, Luxembourg (e-mail: tapapazaf@gmail.com). S. Chatzinotas is with the SnT at the University of Luxembourg, Luxembourg (e-mail: symeon.chatzinotas@uni.lu). M. Debbah is with KU 6G Research Center, Khalifa University of Science and Technology, Abu Dhabi 127788, UAE (e-mail: merouane.debbah@ku.ac.ae).}
}

\markboth{Draft}%
{Shell \MakeLowercase{\textit{et al.}}:Stacked Intelligent Metasurfaces for }

\maketitle

\begin{abstract}
Stacked intelligent metasurfaces (SIM) have recently emerged as a promising technology, which can realize transmit precoding in the wave domain. In this paper, we investigate a SIM-aided integrated sensing and communications system, in which SIM is capable of generating a desired beam pattern for simultaneously communicating with multiple downlink users and detecting a radar target. Specifically, we formulate an optimization problem of maximizing the spectrum efficiency, while satisfying the power constraint of the desired direction. This requires jointly designing the phase shifts of the SIM and the power allocation at the base station. By incorporating the sensing power constraint into the objective functions as a penalty term, we further simplify the optimization problem and solve it by customizing an efficient gradient ascent algorithm. Finally, extensive numerical results demonstrate the effectiveness of the proposed wave-domain precoder for automatically mitigating the inter-user interference and generating a desired beampattern for the sensing task, as multiple separate data streams transmit through the SIM.
\end{abstract}


\vspace{-0.5cm}
\section{Introduction}
\IEEEPARstart{W}{ith} the rapid development of emerging intelligent services such as intelligent transportation, drones, and the Internet of Things, next-generation wireless networks not only need to support stringent communication performance such as high transmission rate and low latency, but also to provide high-accuracy sensing service, such as detection, localization, and tracking\cite{ref10}. Therefore, integrated sensing and communications (ISAC) is considered to be one of the core features of the sixth-generation wireless network.

In ISAC systems, radar and communication devices need to share the same hardware and waveform. Additionally, the constraints introduced by sensing capabilities will limit the degrees of freedom in waveform design\cite{ref10}. To solve this, significant efforts have been made focusing on the unified beamforming design \cite{ref7,ref9,wang2022partially,ref20,ref23}. While these ISAC schemes achieve favorable performance trade-offs between sensing and communications, they generally rely on digital beamforming. When an extremely large aperture array is utilized, the conventional fully digital architecture typically requires a large number of radio frequency (RF) chains, resulting in high hardware costs. Although hybrid beamforming schemes could lower the hardware cost, the constant-modulus constraint of the analog component reduces the degrees of freedom in waveform design, which would lead to a performance penalty compared to the digital schemes assigning each antenna with an RF chain. To address this challenge, stacked intelligent metasurfaces (SIM) have been proposed to enable wave-based beamforming, which can achieve full-precision digital beamforming while reducing the hardware cost\cite{ref1}. Thus, SIM has the potential to replace traditional digital beamforming \cite{ref17}.

Specifically, SIM consists of multiple layers of programmable metasurfaces with a structure similar to an artificial neural network \cite{arXiv_2024_Yang_Joint, arXiv_2024_Wang_Multi, arXiv_2024_Liu_Multi}. By appropriately designing its hardware structure and optimizing the transmission coefficients of meta-atoms, SIM can carry out advanced signal processing directly in the native electromagnetic (EM) wave regime, such as matrix operations\cite{NE_2022_Liu_A,ref1}. This is in contrast to conventional single-layer reconfigurable intelligent surfaces (RIS), which are typically utilized to reshape wireless propagation environments \cite{ref18,ref20}.

To elaborate, {\it{An et al.}} \cite{ref1} proposed a novel SIM-aided holographic MIMO framework, which accomplishes transmit precoding and reception combining automatically as the EM waves pass through the SIM. To further exploit the potential of SIM, the authors of \cite{ref3} applied SIM to the downlink of a multi-user MISO system where the interference between user equipments (UEs) is efficiently suppressed. Following this, substantial works examine the application of the SIM in practical wireless systems, such as the SIM-aided channel modeling \cite{ref19} and estimation\cite{ref11}, hybrid digital and wave-based architecture\cite{ref12}, etc. It was demonstrated that by leveraging multiple metasurface layers, SIM possesses powerful computing power and outperforms its single-layer counterparts \cite{ref1,ref3, arXiv_2024_Huang_Stacked, ref14,ref15,ref18, arXiv_2024_Liu_Stacked, TCOM_2024_Li_Stacked}. Besides, by modeling the input layer of SIM as a uniform planar array (UPA) and training its multilayer architecture, SIM can perform the two-dimensional discrete Fourier transform and estimate the direction-of-arrival of the target \cite{ref13}. Based on these observations, SIM has the potential to generate dual-functional beampatterns for ISAC applications, which, however, has not been hitherto studied yet.

Against this background, we present a SIM-aided ISAC system in this paper, focusing particularly on maximizing the spectrum efficiency (SE) of downlink communication systems while directing a beam towards a single target of interest. In contrast to the conventional digital precoding schemes, a SIM is capable of generating multiple beams towards a radar target and communication users with minimized interference, so as to achieve the goal of dual-functional integration of communication and sensing relying on wave-based computation.

{\it{Notations}}: Bold lowercase and uppercase letters denote vectors and matrices, respectively; ${\left( \cdot \right)^H}$ represents the Hermitian transpose; $\Im\left( c \right)$ denotes the imaginary part of a complex number $c$; $\left\| \cdot \right\|_{\text{F}}$ is the Frobenius norm; $\mathbb{E}\left( \cdot \right)$ stands for the expectation operation; $\log_{u} \left( \cdot \right)$ is the logarithmic function with base $u$; $\textrm{diag}\left ( \mathbf{v} \right )$ produces a diagonal matrix with the elements of $\mathbf{v}$ on the main diagonal; $\textrm{sinc}\left ( x \right )=\sin\left ( \pi x \right )/\left (\pi x \right )$ is the sinc function; ${{\mathbb{C}}^{x \times y}}$ represents the space of $x \times y$ complex-valued matrices; The distribution of a circularly symmetric complex Gaussian random vector with a mean vector $\bold{v}$ and a covariance matrix $\bold{\Sigma}$ is expressed as  $\sim\mathcal{CN}(\bold{v},\bold{\Sigma})$, where $\sim$ stands for “distributed as”; ${{\bf{I}}_N} \in \mathbb{C}^{N \times N}$ denotes the identity matrix. \IEEEpubidadjcol

\section{System Model}
\subsection{SIM Model}
Fig. 1 illustrates the proposed ISAC system, where a SIM serves as the radome of the base station (BS), facilitating the shared use of all BS antennas for both downlink communication to K UEs and radar sensing of a target of interest. Additionally, we highlight that in practice the communication signal is generally known by the transmitter and collaborative receiver. Hence, the communication signal is also employed as the radar probing waveform to enable dual function\cite{ref7}. Note that in contrast to conventional systems, SIM carries out ISAC beamforming in the wave domain for efficiently eliminating the multi-user interference and generating a desired beam.

The SIM consists of $Q$ evenly distributed metasurface layers, with each layer comprising $M$ meta-atoms\cite{ref1}. These meta-atoms can manipulate the EM behavior, as the waves pass through them. In the proposed system, each data stream is sent separately from a different antenna to the UE, and beamforming is performed automatically as information-carrying EM waves propagate through the SIM, eliminating the need for the digital beamformer. However, the number of antennas typically differs from the number of UEs, which needs antenna selection in practice\cite{ref14}. Since we are focusing on wave-domain beamforming for ISAC, we assume that both the number of antennas and the number of users are $K$ for simplicity. Let $\mathcal{Q}=\{1,2,\ldots,Q\},\ \mathcal{M}=\{1,2,\ldots,M\},\ \mathcal{K}=\{1,2,\ldots,K\}$ represent the sets of metasurfaces, meta-atoms on each layer, and UEs, respectively. The diagonal phase shift matrix of the $q$-th metasurface layer can be written as
\begin{equation}\label{eq1} \bold{\Phi}_q=\textrm{diag}(e^{j\Omega^q_1},e^{j\Omega^q_2},\ldots,e^{j\Omega^q_M})\in \mathbb{C}^{M\times M},q\in\mathcal{Q},
\end{equation}
where $e^{j\Omega^q_m},\forall{q}\in\mathcal{Q},\forall{m}\in\mathcal{M}$ 
denotes the EM response of the $m$-th meta-atom on the $q$-th layer, and $\Omega^q_m\in [0,2\pi)$ is the corresponding phase shift. We can connect these meta-atoms to a smart controller and produce a customized spatial waveform shape at the output of the metasurface layer by subtly configuring the SIM phase shifts\cite{ref2, NE_2022_Liu_A}.

Moreover, the transmission matrix from the $(q-1)$-st to the $q$-th metasurface layer can be written as $\bold{W}_q\in \mathbb{C}^{M\times M},\forall{q}\ne 1,q\in \mathcal{Q}$. According to the Rayleigh-Sommerfeld diffraction theory, the $(m,m')$-th entry $w^q_{m,m'}$ of $\bold{W}_q$ is given by $w^q_{m,m'}=\frac{A_t\cos\psi^q_{m,m'}}{d^q_{m,m'}}\Bigl(\frac{1}{2\pi d^q_{m,m'}}-j\frac{1}{\lambda}\Bigr)e^{j2\pi d^q_{m,m'}/\lambda}$, where $\lambda$ is the wavelength, $A_t$ is the area of each meta-atom in the SIM, $\psi^q_{m,m'}$ represents the angle between the propagation direction and the normal direction of the $(q-1)$-st transmit metasurface layer, while $d^q_{m,m'}$ denotes the corresponding propagation distance. In particular, we define $\bold{W}_1=[\bold{w}^1_1,\bold{w}^1_2,\ldots,\bold{w}^1_K]\in \mathbb{C}^{M\times K}$, and $\bold{w}^1_k\in \mathbb{C}^{M\times 1},k\in\mathcal{K}$ represents the transmission vector from the $k$-th antenna to the first metasurface layer of the SIM. Similarly, the \(m\)-th entry \( w^1_{m,k} \) of \( \mathbf{w}^1_k \) is obtained by replacing $\psi^q_{m,m'}$ and $d_{m,m'}$ in the preceding expression with $\psi^1_{m,k}$ and $d^1_{m,k}$, respectively. As a result, the transfer function of SIM is formulated by \cite{ref1,ref2,ref3,Science_2018_Lin_All}
\begin{equation}\label{eq3} 
 \bold{G} = \bold{\Phi}_Q \bold{W}_Q \bold{\Phi}_{Q-1} \bold{W}_{Q-1} \ldots \bold{\Phi}_2 \bold{W}_2 \bold{\Phi}_1\in \mathbb{C}^{M\times M}.
\end{equation}

\begin{figure}[!t]
\centering
\includegraphics[width=0.7\textwidth,interpolate=true]{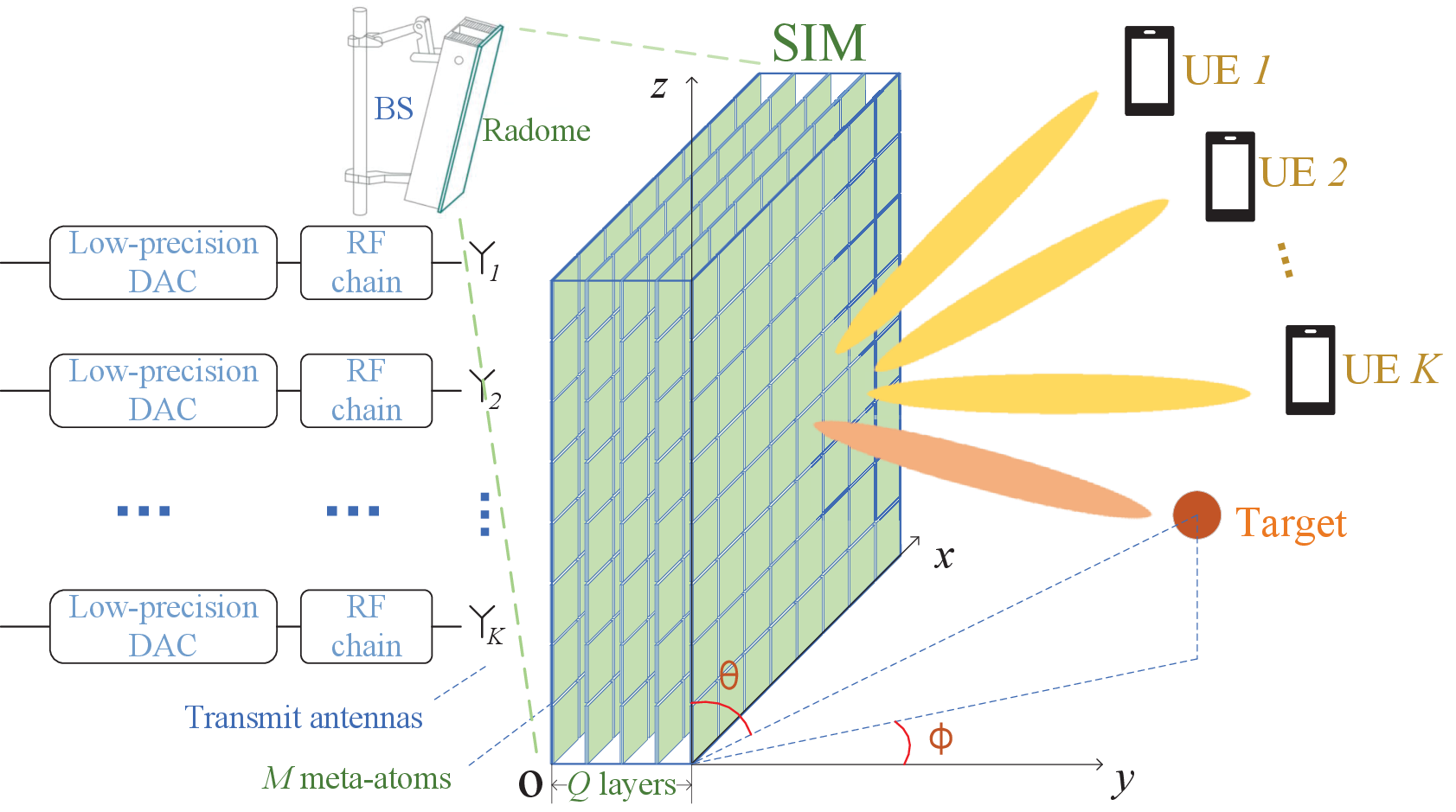}
\caption{{A SIM-aided ISAC system.}}\vspace{-0.5cm}
\label{fig_1}
\end{figure}

\vspace{-0.45cm}

\subsection{Signal Model}
We denote $\bold{s}\in \mathbb{C}^{K\times 1}$, where $s_k\sim\mathcal{CN}(0,1),k\in\mathcal{K}$, i.e., the $k$-th entry of $\bold{s}$, represents the information symbol intended for UE $k$, and assume that they are mutually independent such that $\mathbb{E}\left( \bold{ss}^H \right) = \bold{I}_K$\cite{ref7}.
The transmission signal vector from the output layer of the SIM is given by
\begin{equation}\label{eq4} 
\bold{t}=\bold{G}\bold{W}_1\bold{P}\bold{s}\in \mathbb{C}^{M\times 1},
\end{equation}
where $\bold{GW}_1\in \mathbb{C}^{M\times K}$ characterizes the transfer function from the antenna array to the output layer of the SIM, $\bold{P}=\textrm{diag}(\sqrt{p_1},\sqrt{p_2},\ldots,\sqrt{p_K})\in \mathbb{C}^{K\times K}$ represents the power allocation for $K$ UEs. Thus, the waveform covariance matrix is written as $\bold{R_t}=\mathbb{E}\left( \bold{tt}^H \right)=\bold{GW}_1\bold{PP}^H\bold{W}^H_1\bold{G}^H\in \mathbb{C}^{M\times M}$.
\vspace{-0.1cm}
\subsection{Communication Model}
We define $\bold{H}=[\bold{h}_1,\bold{h}_2,\ldots,\bold{h}_K]^H\in \mathbb{C}^{K\times M}$, where $\bold{h}^H_k \in \mathbb{C}^{1\times M},k\in\mathcal{K}$ is the propagation channel from the SIM to the $k$-th user \cite{ref3}.
We assume that the channel $\bold{H}$ between the SIM and $K$ UEs is flat Rayleigh fading, i.e., 
\begin{equation}\label{eq7}
 \bold{h}^H_k\sim\mathcal{CN}(\bold{0},\upsilon_k\bold{R}_Q),k\in\mathcal{K},
\end{equation}
where $\bold{h}^H_k$ has been perfectly estimated by utilizing uplink pilot signals \cite{ref12}, and $\upsilon_k$ represents the distance-dependent path loss for the communication link between SIM and the $k$-th UE. $\upsilon_k$ is modeled as $\upsilon_k = C_0(D_k/D_0)^{-\bar{n}}$,
where $D_k$ denotes the propagation distance between the SIM and the $k$-th UE. Additionally, the free space path loss at a reference distance of $D_0 =1$m is represented by $C_0=(\lambda/4\pi D_0)^2$, and $\bar{n}$ represents the path loss exponent. Additionally, $\bold{R}_Q\in \mathbb{C}^{M\times M}$ quantifies the spatial correlation among the channels associated with different meta-atoms on the final layer. When considering an isotropic scattering environment with uniformly distributed multipath components, $\bold{R}_Q$ is expressed by $[\bold{R}_Q]_{m,m'}=\textrm{sinc}\Bigl({2E_{m,m'}}/{\lambda}\Bigr)$, where $E_{m,m'}$ denotes the corresponding meta-atom spacing\cite{ref21}. Thus, the signal received by the $k$-th user can be expressed as
\begin{equation}\label{eq10} 
 y_k=\bold{h}^H_k\bold{G}\sum\nolimits_{k'=1}^K\bold{w}_{k'}^1\sqrt{p_{k'}}s_{k'}+n_k,k\in\mathcal{K},
\end{equation}
where $n_k\sim\mathcal{CN}(0,N_k)$ represents additive white Gaussian noise (AWGN) and $N_k$ is the average noise power at UE $k$.

\subsection{Sensing Model}
Specifically, the output layer of the SIM is modeled as a UPA with antenna spacing of $\lambda/2$. By properly configuring the phase shifts of meta-atoms in each layer, the SIM can emit a concentrated beam toward the target's direction. The channel between the SIM and the radar’s targets is modeled as a line-of-sight propagation channel\cite{ref7}. As shown in Fig. \ref{fig_1}, we consider a far-field target with the azimuth angle of $\phi$ and the elevation angle of $\theta$. $M_x$ and $M_z$ denote the number of meta-atoms in the $x$-axis and $z$-axis, respectively. The steering vector of the SIM is derived as\cite{ref13}
\begin{equation}\label{eq13} 
 \bold{a}(\theta,\phi)=\frac{1}{\sqrt{M_xM_z}}\bold{a}_x(\theta,\phi)\otimes \bold{a}_z(\theta)\in \mathbb{C}^{M\times 1},
\end{equation}
where we have $\theta\in(0,\pi),\phi\in(-\pi/2,\pi/2)$. Besides, $\bold{a}_x(\theta,\phi)=[1,e^{-j\pi \sin\theta \sin\phi},\ldots,e^{-j(M_x-1)\pi \sin\theta \sin\phi}]^T\in \mathbb{C}^{M_x\times1}$ and $\bold{a}_z(\theta)=[1,e^{-j\pi \cos\theta},\ldots,e^{-j(M_z-1)\pi \cos\theta}]^T\in \mathbb{C}^{M_z\times1}$ are the steering vector of $x$-axis and $z$-axis, respectively.

Assuming that the radar target is in the direction of $(\theta_c,\phi_c)$, the beampattern gain directed towards the target is written as $P(\theta_c,\phi_c)=\bold{a}(\theta_c,\phi_c)^H\bold{R_t}\bold{a}(\theta_c,\phi_c)$\cite{ref9}.

In this paper, we aim to design the SIM response to make sure that $P(\theta_c,\phi_c)$ meets the desired value for the sensing task's requirement.

\section{Problem Formulation and Solution}
\subsection{Problem Formulation}
In this subsection, we formulate an optimization problem of maximizing the SE by jointly optimizing the power allocation coefficients at the BS and the phase shifts of the SIM, subject to the total power $P_t$ and desired beampattern constraints. Specifically, the optimization problem is formulated as
\begin{subequations}\label{eq17}
 \begin{align}
\mathop{\textrm{max}}\limits_{\bold{\Phi}_q,\bold{P}}\quad&R=\sum\nolimits_{k=1}^K\log_2(1+\gamma_k)\\
\text{s.t.}\quad&\bold{a}^H(\theta_c,\phi_c)\bold{R_t}\bold{a}(\theta_c,\phi_c)\ge\Psi,\\
&\bold{R_t}=\bold{GW}_1\bold{PP}^H\bold{W}^H_1\bold{G}^H,\\
&\gamma_k=\frac{|\bold{h}^H_k\bold{G}\bold{w}^1_k|^2p_k}{\sum\nolimits_{k'\ne k}^K|\bold{h}^H_{k}\bold{G}\bold{w}^1_{k'}|^2p_{k'}+N_k},\forall{k}\in\mathcal{K},\\
&\sum\nolimits_{k=1}^Kp_k\le P_t,\\
&p_k\ge 0,\forall{k}\in\mathcal{K},\\
&(\ref{eq1})-(\ref{eq3})
\end{align}
\end{subequations}
where $\Psi$ denotes the desired beampattern gain at the target's direction, and $\gamma_k$ is the signal-to-interference-plus-noise ratio of UE $k$. In order to simplify the problem in (\ref{eq17}), we transform the beamforming pattern constraint as a penalty term to the objective function. Then the problem can be reformulated as follows:
\begin{subequations}\label{eq18}
\begin{align}
\mathop{\textrm{max}}\limits_{\bold{\Phi}_q,\bold{P}}\quad&F=\sum\nolimits_{k=1}^K\log(1+\gamma_k)+\beta g_{c}\\
\text{s.t.}\quad&g_{c}=\textrm{min}\{ \bold{a}^H(\theta_c,\phi_c)\bold{R_t}\bold{a}(\theta_c,\phi_c)-\Psi,0\},\\
&(\ref{eq17}\textrm{c})-(\ref{eq17}\textrm{g}),
\end{align}
\end{subequations}
where $\beta\in \mathbb{R^{+}}$ in the second term represents the penalty factor for striking a flexible tradeoff between the sum rate and the desired beampattern gain. Then, the optimal solution of the constrained problem (\ref{eq17}) is approached by solving problem (\ref{eq18}). 

\subsection{The Proposed Gradient Ascent Algorithm}
To address problem \eqref{eq18}, we propose an efficient gradient ascent algorithm to iteratively adjust the phase shifts of the SIM and the power allocation coefficients until reaching convergence. The specific steps of the algorithm are outlined as follows:

\textit{\textbf{Step 1: Initialize the phase shifts and power allocation coefficients}}

The phase shifts of the SIM are randomly initialized and the power allocation coefficients are obtained using the water-filling algorithm. In order to avoid the gradient ascent method falling into a local optimal solution, we first generate multiple sets of phase shifts and then select the SIM configuration resulting in the maximum value of $F$ for initialization\cite{ref1}. 

\textit{\textbf{Step 2: Calculate the partial derivatives}}

For $\forall{q}\in\mathcal{Q},\forall{m}\in\mathcal{M}$, the gradient of the objective function $F$ with respect to the phase shift of the $m$-th meta-atom on the $q$-th layer of the SIM is calculated by
\begin{equation}\label{eq19} 
 \frac{\partial{F}}{\partial{\Omega^q_m}}=\frac{\partial{R}}{\partial{\Omega^q_m}}+\beta\frac{\partial{g_c}}{\partial{\Omega^q_m}}.
\end{equation}
In order to simplify the expression, we firstly define $\delta_{k,k'}=|\bold{h}_k^H\bold{Gw}_{k'}^1|^2p_{k'}$ and write $\bold{a}(\theta_c,\phi_c)$ as $\bold{a}$. Hence, the two terms of \eqref{eq19} can be explicitly expressed as 
\begin{equation}\label{eq20} 
\begin{split}
 \frac{\partial{R}}{\partial{\Omega^q_m}}=\log_2e\sum\limits_{k=1}^K\frac{1}{1+\gamma_k}\bigl(\frac{\gamma_k}{\delta_{k,k}}\frac{\partial\delta_{k,k}}{\partial\Omega^q_m}-\frac{\gamma_k^2}{\delta_{k,k}}\sum\limits_{k'\ne k}^K\frac{\partial\delta_{k,k'}}{\partial\Omega^q_m}\bigr),
\end{split}
\end{equation}
\vspace{-0.3cm}
\begin{equation}\label{eq21} 
\begin{split}
 \frac{\partial{g_c}}{\partial{\Omega^q_m}}=\sum\limits_{k=1}^Kp_k\Im\left[ (e^{j\Omega^q_m}\bold{aL}_{:,m}\bold{K}_{m,:}\bold{w}_k^1)^H(\bold{aGw}_k^1) \right],
\end{split}
\end{equation}
where $\frac{\partial\delta_{k,k'}}{\partial\Omega^q_m}$ is obtained by
\begin{equation}\label{eq22} 
\begin{split}
 \frac{\partial\delta_{k,k'}}{\partial\Omega^q_m}=p_{k'}\Im\left[ (e^{j\Omega^q_m}\bold{h}_k^H\bold{L}_{:,m}\bold{K}_{m,:}\bold{w}_{k'}^1)^H(\bold{h}_k^H\bold{Gw}_{k'}^1) \right].
\end{split}
\end{equation}

Besides, $\bold{K}^q_{m,:}$ and $\bold{L}^q_{:,m}$ denote the $m$-th row of $\bold{K}^q\in\mathbb{C}^{M\times M}$ and the $m$-th column of $\bold{L}^q\in\mathbb{C}^{M\times M}$ respectively. And $\bold{K}^q$ and $\bold{L}^q$ are defined as
\begin{equation}\label{eq23} 
 \bold{K}^q=
 \begin{cases}
 \bold{W}_q\bold{\Phi}_{q-1}\bold{W}_{q-1}\ldots\bold{\Phi}_2\bold{W}_2\bold{\Phi}_1,&q\ne1,\\
 \bold{I}_M,&q=1,
 \end{cases}
\end{equation}
\begin{equation}\label{eq24} 
 \bold{L}^q=
 \begin{cases}
 \bold{\Phi}_Q\bold{W}_Q\bold{\Phi}_{Q-1}\ldots\bold{\Phi}_{q+1}\bold{W}_{q+1},&q\ne Q,\\
 \bold{I}_M,&q=Q.
 \end{cases}
\end{equation}

\begin{algorithm}[t]
\caption{The proposed gradient ascent algorithm for 
solving problem \eqref{eq18}}\label{alg:alg1}
\begin{algorithmic}
\STATE 
\STATE 1: {\textbf{Input:}} $\mathbf{H},\mathbf{W}_q,\forall{q}\in\mathcal{Q},P_t,\beta,\Psi$;
\STATE 2: Initialize $\bold{\Phi}_q,\forall{q}\in\mathcal{Q}$ randomly and initialize $\bold{P}$ using the water-filling algorithm;
\STATE 3: {\textbf{Repeat}}
\STATE 4: Calculate the partial derivatives of $F$ with respect to $\Omega^q_m$ and $p_k$ by applying \eqref{eq19} and \eqref{eq25}, respectively;
\STATE 5: Normalize the partial derivatives by applying \eqref{eq28} and \eqref{eq29}, respectively;
\STATE 6: Update $\Omega^q_m$ and $p_k$ by applying \eqref{eq30} and \eqref{eq31}, respectively; 
\STATE 7: {\textbf{Until}} The objective function $F$ converges;
\STATE 8: {\textbf{Output:}} $\bold{P},\bold{\Phi}_q,\forall{q}\in\mathcal{Q}$.
\end{algorithmic}
\label{alg1}
\end{algorithm}

Next, the gradient of the objective function $F$ with respect to the power allocation coefficients at the BS is calculated by
\begin{align}\label{eq25}
 \frac{\partial{F}}{\partial{p_k}}&=\frac{\partial{R}}{\partial{p_k}}+\beta\frac{\partial{g_c}}{\partial{p_k}},\\
 \frac{\partial{R}}{\partial{p_k}}&=\log_2 e[\frac{\gamma_k}{(1+\gamma_k)p_k}-\sum_{k'\ne k}^K\frac{\gamma_{k'}^2\delta_{k',k}}{\delta_{k',k'}(1+\gamma_{k'})p_k}],\\
 \frac{\partial{g_c}}{\partial{p_k}}&= \bold{aGw}_k^1{(\bold{aGw}_k^1)}^H.
\end{align}

\textit{\textbf{Step 3: Normalize the partial derivatives}}

To reduce the risk of oscillation during optimization, we normalize the calculated partial derivatives as follow\cite{ref1}
\begin{align}
 \frac{\partial F}{\partial\Omega^q_m}&\gets\frac{\pi}{\eta}\frac{\partial F}{\partial\Omega^q_m},\forall{m}\in\mathcal{M},\forall{q}\in\mathcal{Q},\label{eq28}\\
 \frac{\partial F}{\partial p_k}&\gets\frac{P_t}{\rho K}\frac{\partial F}{\partial p_k},\forall{k}\in\mathcal{K},\label{eq29}
\end{align}
where $\eta=\textrm{max}(\frac{\partial F}{\partial\Omega^q_m}),\forall{m}\in\mathcal{M},\forall{q}\in\mathcal{Q}$ and $\rho=\textrm{max}(\frac{\partial F}{\partial p_k}),\forall{k}\in\mathcal{K}$ represent, respectively, the maximum value of the partial derivative with respect to the phase shifts of the SIM and the maximum value of the partial derivative with respect to the power allocation coefficients.

\textit{\textbf{Step 4: Update the phase shifts and power allocation coefficients}}

Then, we can update the phase shifts of the SIM and the power allocation coefficients by
\begin{align}
 \Omega^q_m&\gets\Omega^q_m+\mu\frac{\partial F}{\partial\Omega^q_m},\forall{m}\in\mathcal{M},\forall{q}\in\mathcal{Q},\label{eq30}\\
 p_k&\gets p_k+\mu\frac{\partial F}{\partial p_k},\forall{k}\in\mathcal{K},\label{eq31} 
\end{align}
where $\mu>0$ represents the Armijo step size, which is determined by utilizing the backtracking line search during each iteration \cite{ref3}. In order to satisfy (\ref{eq17}h) and (\ref{eq17}i), we first compensate the power of all UEs to make sure that the minimum value is larger than zero in every update. Then we perform normalization by taking
$p_k=p_k\cdot{P_t}/{(\sum_{k=1}^Kp_k)}$.

\begin{figure}[t]
\vspace{-0.4cm}
 \centering
 \begin{minipage}[b]{0.49\linewidth}
 \includegraphics[width=\textwidth]{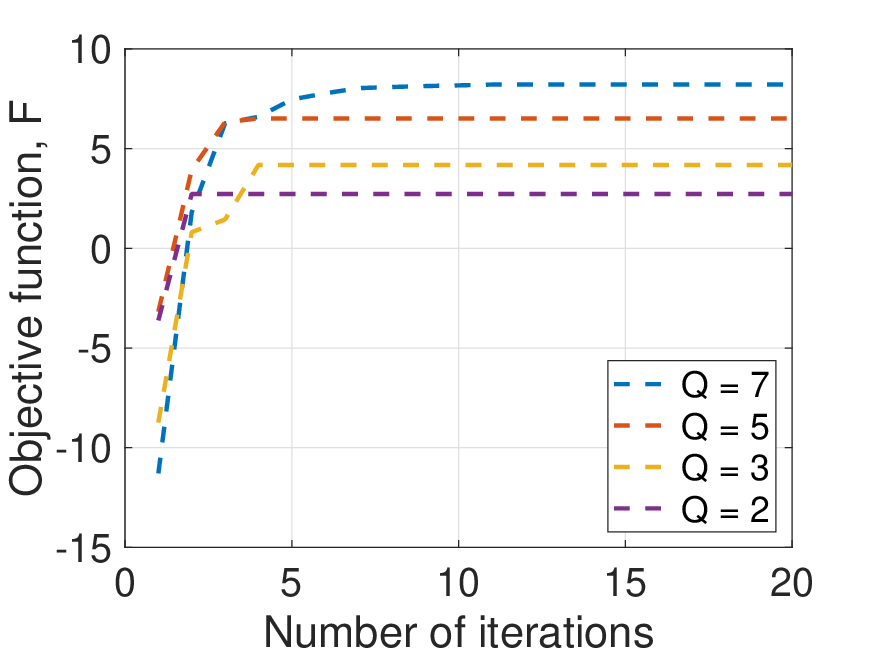}
 \caption{The convergence curves of the proposed gradient ascent algorithm ($M=100,\Gamma = 8 ~\text{dBi}$).}
 \label{fig:image1}
 \end{minipage}
 \hfill
 \begin{minipage}[b]{0.49\linewidth}
 \includegraphics[width=\textwidth]{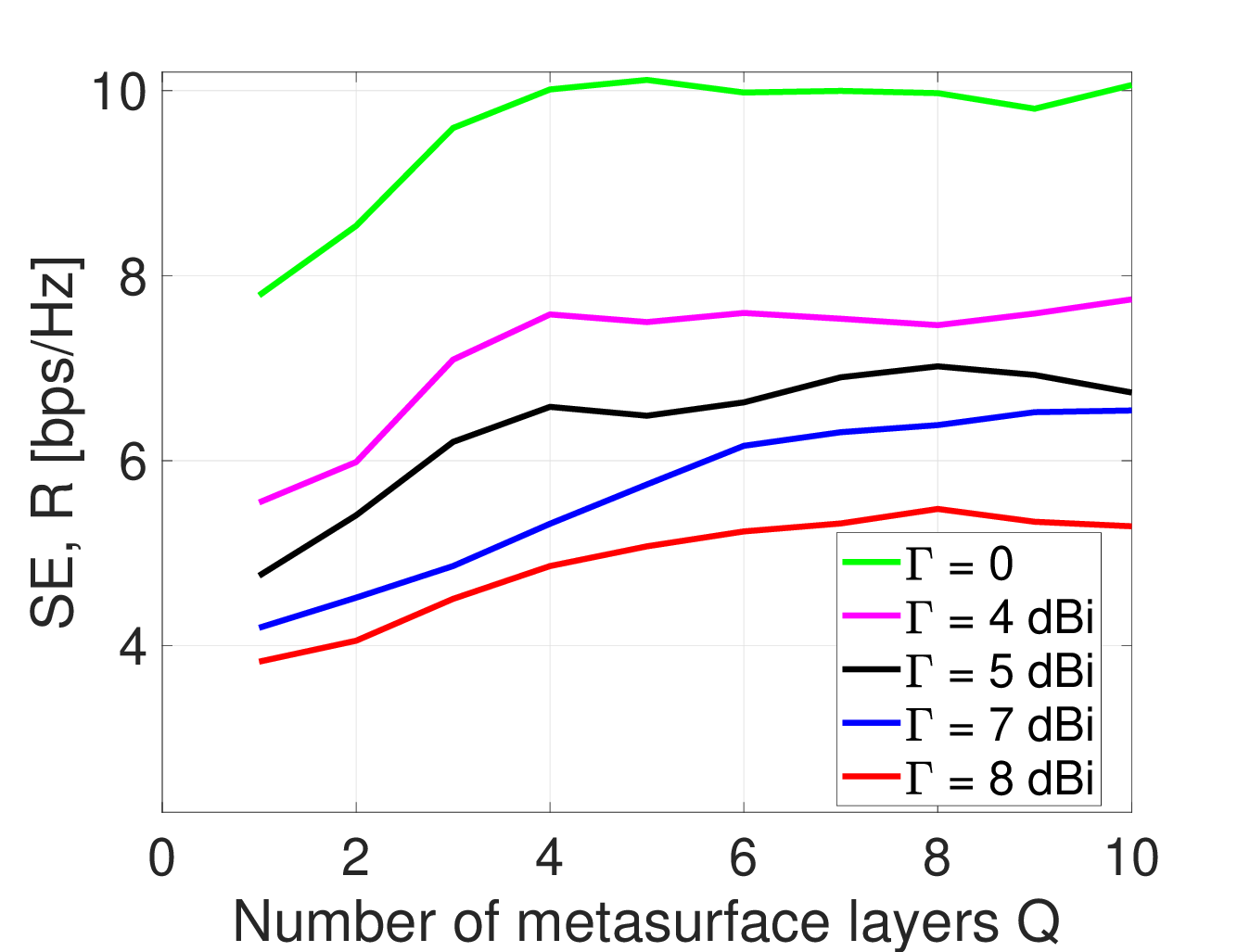}
 \caption{The SE versus the number of metasurface layers ($M=100$).}
 \label{fig:image2}
 \end{minipage}
 \vspace{-0.5cm}
\end{figure}

By repeating \textit{\textbf{Steps 2}}-\textit{\textbf{4}}, we obtain the phase shifts of the SIM and power allocation coefficients as the objective function value $F$ converges. For clarity, we summarize the major steps of the proposed gradient ascent method in Algorithm 1.

\section{Numerical Results}
In this section, we provide numerical results for characterizing the performance of the proposed SIM-aided ISAC system.
\vspace{-0.8cm}
\subsection{Simulation Setups}
In the SIM-aided ISAC system shown in Fig. 1, choosing the appropriate hardware parameters for SIM is essential. For instance, having too many meta-atoms and layers can increase the computational complexity of solving the problem, whereas having too few may result in insufficient performance\cite{ref1,ref18}. Thus, we set $M=100, K=4, P_t=15 ~\textrm{dBm} ~\textrm{and}~N_k = -104~\textrm{dBm}$. The total thickness of the SIM is $5\lambda$, so the distance between adjacent metasurface layers is $5\lambda/Q$. The spacing between meta-atoms is $\lambda/2$.
To make sure a fair comparison, we set $\Psi=\alpha \Gamma$, where $\Gamma$ is the normalized beampattern gain and $\alpha={\left\| \bold{R_t} \right\|_{\text{F}}}/{\sqrt{M}}$ is the scale factor. The beampattern gain is evaluated with respect to the corresponding omnidirectional beampattern gain. The coordinates of the users are $(10k, 10, 10k), k = 1,2,3,4$, and the channel between the SIM and UE $k$ is described by \eqref{eq7}, where we have $\bar{n}=3.5$. In addition, we consider a system that operates at a carrier frequency of 28 GHz. The target of interest is located at $(90^\circ,45^\circ)$. The penalty factor $\beta$ is set to 2.

\vspace{-0.4cm}
\subsection{Performance Evaluation of the Proposed Algorithm}

Fig. 2 verifies the convergence performance of the proposed gradient ascent algorithm for $M=100, \Gamma=8~\textrm{dBi}$, from which we can see that the objective function reaches convergence within 10 iterations under different setups. Moreover, a larger number of metasurface layers results in a higher SE, thanks to the array gain provided by a large metasurface aperture.

In Fig. 3, we compare the SE under different beampattern gain thresholds $\Gamma$. Observe from Fig. 3 that the SE gradually decreases as $\Gamma$ increases. This is due to the fact that as $\Gamma$ increases, the increased power allocation for sensing tasks results in reduced power for UEs and constraints on waveform design. Besides, in comparison to a single-layer configuration, a 7-layer SIM exhibits a remarkable 33\% SE gain, as the multi-layer architecture of the SIM provides more powerful computing capability to mitigate inter-user interference in the wave domain. Nevertheless, when the number of metasurface layers exceeds a certain value, overly dense SIM layers may cause performance degradation. This is because the SIM's transfer function in (3) highly relies on an appropriate inter-layer propagation matrix, while a small layer spacing would result in the transmission matrix $\bold{W}_q,q\in\mathcal{Q}$ becoming diagonal. In practical ISAC systems, one should take into account the fundamental tradeoffs to determine the best SIM hardware parameters first and then configure its phase shifts for realizing desired computing functionality in the wave domain.
\begin{figure}[t]
\vspace{-0.5cm}
 \centering
 \begin{tabular}{@{\extracolsep{\fill}}c@{}c@{\extracolsep{\fill}}}
 \includegraphics[width=0.5\linewidth]{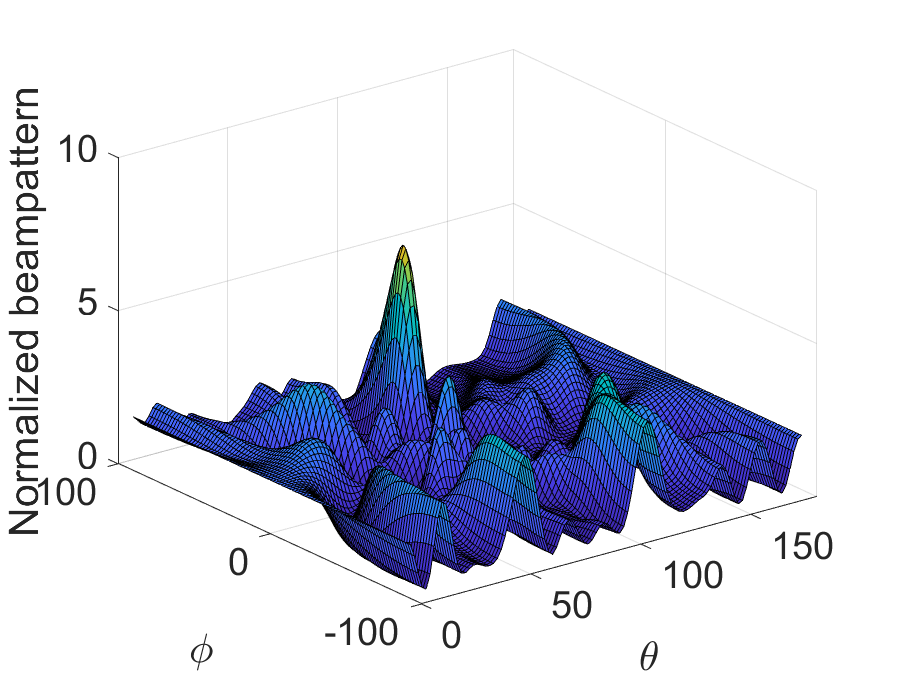} &
 \includegraphics[width=0.5\linewidth]{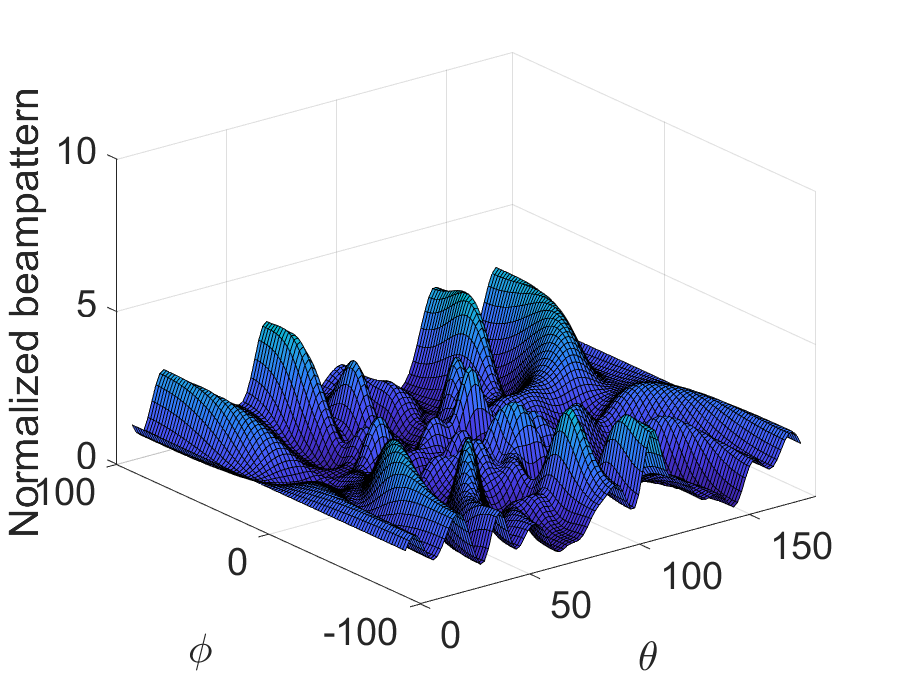}\\
 \resizebox{0.03\linewidth}{!}{(a)}&\resizebox{0.03\linewidth}{!}{(b)}\\
 \end{tabular}
 \caption{Beampattern comparison for $M=100, Q=7, K=4, \Gamma=8~\textrm{dBi}$. (a) ISAC; (b) communication-only system.}
 \label{fig4}
 \end{figure}
\begin{figure}[t]
 \centering
 \begin{tabular}{@{\extracolsep{\fill}}c@{}c@{\extracolsep{\fill}}}
 \includegraphics[width=0.5\linewidth]{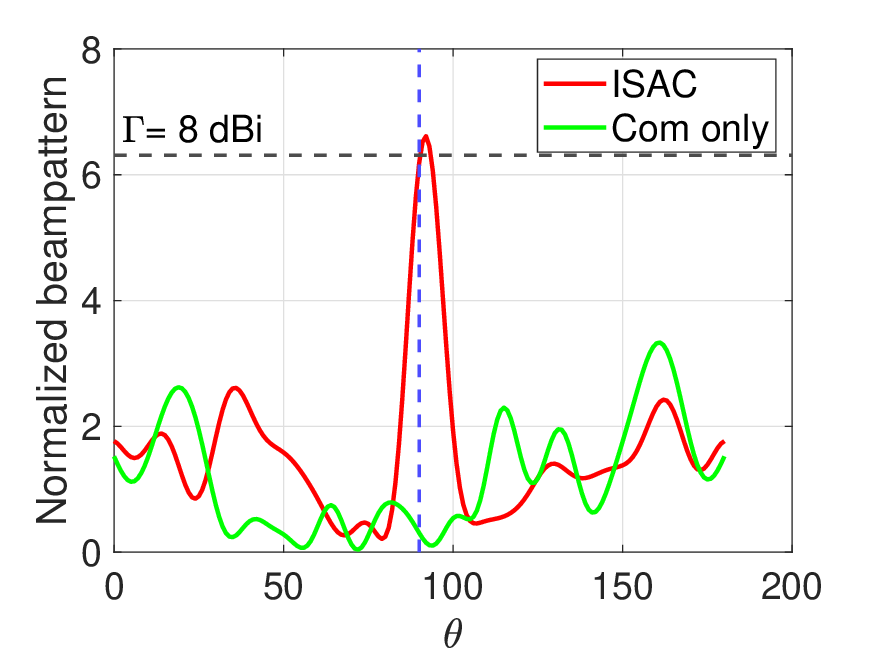} &
 \includegraphics[width=0.5\linewidth]{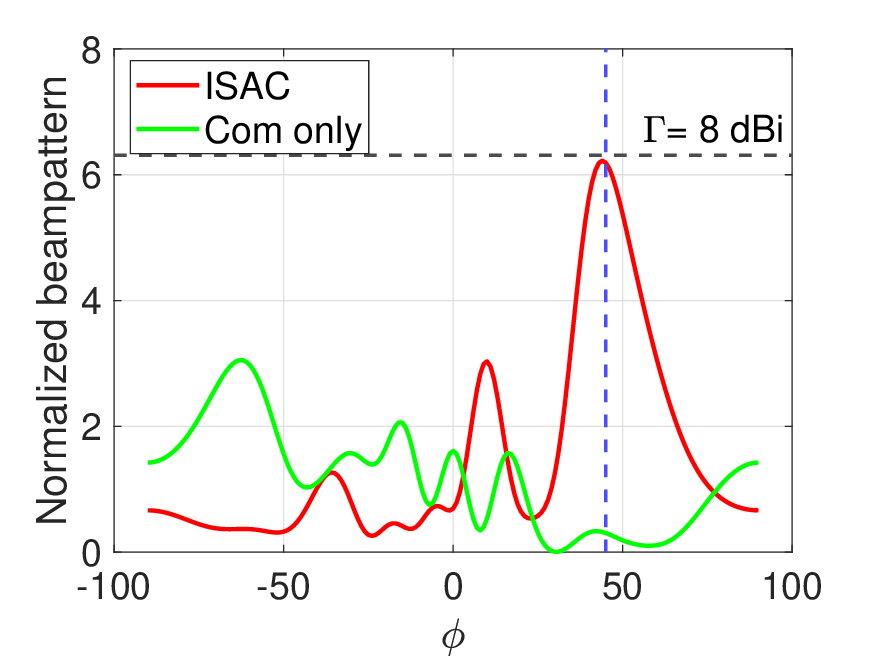}\\
 \resizebox{0.03\linewidth}{!}{(a)}&\resizebox{0.03\linewidth}{!}{(b)}\\
 \end{tabular}
 \caption{Profiles of Fig. 4. (a) Horizontal beampattern; (b) Vertical beampattern.}\vspace{-0.5cm}
 \label{fig5}
 \end{figure}
 
Fig. 4 compares the beampattern generated by the SIM in the ISAC scenario and communication-only system, respectively.
Compared to the communication-only system, the proposed SIM-based transmitting array is capable of generating a desired beam in the target direction by leveraging the communication signals with waveforms known by the local sensing receiver. To illustrate, we also plot horizontal and vertical profiles of SIM-generated 3D beampattern in Figs. 5(a) and (b), respectively. Note that under both setups the beampattern gain in the desired direction satisfies the constraint condition, which demonstrates the effectiveness of the dual-functional precoding in the wave domain.

\vspace{-0.25cm}
\section{Conclusions}
In this paper, a SIM-aided ISAC system was proposed, where all antennas are shared for communication and radar sensing. In contrast to conventional digital beamforming, a SIM was utilized to perform downlink beamforming in the wave domain. We formulated an optimization problem to maximize the SE of the communication users and ensure the desired gain at the sensing target. And extensive simulation results corroborated that the proposed system is capable of serving multiple UEs with satisfactory SE, while ensuring the desired beampattern gain. Besides, our simulation results also demonstrated that an appropriate increase in the number of SIM layers can effectively mitigate the inter-user interference automatically as multiple data streams are transmitted through the SIM. In subsequent work, we will focus on the SIM-aided ISAC system, taking into account the processing of echo signals and utilizing SIM to achieve wave-domain signal processing in a range of sensing functions such as detection, localization, and tracking.

\vspace{-0.2cm}

\bibliographystyle{IEEEtran}
\bibliography{Mybib}
\newpage
\end{document}